\documentclass[conference]{IEEEtran}
\IEEEoverridecommandlockouts
\usepackage{cite}
\usepackage{amsmath,amssymb,amsfonts}
\usepackage{algorithmic}
\usepackage{graphicx}
\usepackage{textcomp}
\usepackage{xcolor}
\usepackage{kbordermatrix}
\usepackage{empheq}
\usepackage[shortlabels]{enumitem}
\usepackage[hidelinks]{hyperref}

\newcommand{\arrayprod}{\mathbin{{\otimes}.{\oplus}}}
\newcommand{\coloneq}{\mathrel{\raisebox{.35pt}{$\colon$}\!\!{=}}}
\newcommand{\se}[1]{#1}
\newcommand{\textdef}[1]{\emph{#1}}
\newcommand{\ma}[1]{\mathbf{#1}}
\newcommand{\concat}{{}^\frown}
\newcommand{\matmul}{\mathbin{@}}

\renewcommand{\_}{\underline{\;\,}}

\setlist[description]{font=\normalfont\itshape}

\makeatletter
\def\footnoterule{\relax%
  \kern-5pt
  \hbox to \columnwidth{\hfill\vrule width 0.5\columnwidth height 0.4pt\hfill}
  \kern4.6pt}
\makeatother

\renewcommand{\IEEEkeywordsname}{Keywords}

\begin{document}

\title{Python Implementation of the Dynamic Distributed Dimensional Data Model\\
}

\author{\IEEEauthorblockN{Hayden Jananthan${}^1$, Lauren Milechin${}^2$, Michael Jones${}^1$, William Arcand${}^1$, William Bergeron${}^1$, \\
David Bestor${}^1$, Chansup Byun${}^1$, Michael Houle${}^1$, Matthew Hubbell${}^1$, Vijay Gadepally${}^{1, 3}$, \\
Anna Klein${}^1$, Peter Michaleas${}^1$, Guillermo Morales${}^1$, Julie Mullen${}^1$, 
Andrew Prout${}^1$, \\
Albert Reuther${}^1$, Antonio Rosa${}^1$, Siddharth Samsi${}^1$, Charles Yee${}^1$, Jeremy Kepner${}^{1, 3, 4}$}
\IEEEauthorblockA{\textit{MIT Lincoln Laboratory Supercomputing Center${}^1$, MIT EAPS${}^2$, MIT CSAIL${}^3$, MIT Math${}^4$}
% %\textit{name of organization (of Aff.)}\\
% Lexington, MA, U.S.A. \\
% $\{$\href{mailto:hayden.jananthan@ll.mit.edu}{hayden.jananthan},\href{mailto:kepner@ll.mit.edu}{kepner},\href{mailto:vijayg@ll.mit.edu}{vijayg}$\}$@ll.mit.edu
}
}

\maketitle

\IEEEtitleabstractindextext{
\begin{abstract}
	Python has become a standard scientific computing language with fast-growing support of machine learning and data analysis modules, as well as an increasing usage of big data. The Dynamic Distributed Dimensional Data Model (D4M) offers a highly composable, unified data model with strong performance built to handle big data fast and efficiently. In this work we present an implementation of D4M in Python. D4M.py implements all foundational functionality of D4M and includes Accumulo and SQL database support via Graphulo. We describe the mathematical background and motivation, an explanation of the approaches made for its fundamental functions and building blocks, and performance results which compare D4M.py's performance to D4M-MATLAB and D4M.jl.
\end{abstract}

}

% Display Title, Author, Abstract and Keywords
\IEEEpeerreviewmaketitle
\IEEEdisplaynontitleabstractindextext

\section{Introduction}

% Footnote & disclaimer statement
\let\thefootnote\relax\footnotetext{This material is based upon work supported by the Assistant Secretary  of  Defense  for  Research  and  Engineering  under Air  Force  Contract  No.  FA8702-15-D-0001,  National  Science  Foundation  CCF-1533644,  and  United  States  Air  Force Research  Laboratory  and  Air  Force  Artificial  Intelligence Accelerator  Cooperative  Agreement  Number  FA8750-19-2-1000. Any opinions, findings, conclusions or recommendations expressed  in  this  material  are  those  of  the  author(s)  and  do not necessarily reflect the views of the Assistant Secretary of Defense  for  Research  and  Engineering,  the  National  Science Foundation,  or  the United  States  Air  Force.}

As the world continues to become more and more data-driven the amount of data we as a society produce and use grows exponentially. To address this ever-growing Big Data the need for software tools designed to handle such data is ever important. Over the last decade Python has become a standard scientific computing language thanks in part to its ease of access and growing body of popular libraries, including the machine learning libraries TensorFlow/Keras \cite{tensorflow2015,TensorFlow2016,keras2015}, PyTorch \cite{pytorch2019}, and Scikit-learn \cite{scikit-learn2011}, data science libraries like pandas \cite{pandas2010}, and the numerical libraries NumPy \cite{harris2020array} and SciPy \cite{SciPy2020}. 

The D4M technology is another such tool designed to handle big data. D4M (Dynamic Distributed Dimensional Data Model) combines sparse linear algebra, associative arrays, fuzzy algebra, distributed arrays, and triple-store/NoSQL databases to provide an interface to a mathematical representation unifying spreadsheets, database tables, matrices, and graphs/networks \cite{Dynamic2012,D4M2015}. D4M has been leveraged in the identification of pathogens \cite{Dodson2015}, support for high-throughput data ingest of whole-system provenance data \cite{Moyer2016}, and the creation of benchmarks for Accumulo \cite{Kepner2014} and SciDB \cite{Samsi2016}. 

The central data model of D4M is the associative array which generalizes the notions of spreadsheets, databases, matrices, and graphs. Figure~\ref{associative array example} shows an example of an associative array represented in tabular form.

\begin{figure}[htb]
	\begin{empheq}[box=\fbox]{equation*}
		\sf{A} = 
		\kbordermatrix{& \text{{artist}} & \text{{duration}} & \text{{genre}} \\
		\texttt{0294.mp3} & \text{Pink Floyd} & \text{6:53} & \text{rock} \\
		\texttt{1829.mp3} & \text{Samuel Barber} & \text{8:01} & \text{classical} \\
		\texttt{7802.mp3} & \text{Taylor Swift} & \text{10:12} & \text{pop}
		}
	\end{empheq}
	\caption{Tabular representation of an associative array $\sf{A}$.}
	\label{associative array example}
	\label{figure 1}
\end{figure}

D4M was first implemented within MATLAB in 2012 and then Julia in 2016 \cite{Julia2016}. Its Python implementation, D4M.py \cite{D4Mpy}, makes the D4M technology available to the wider scientific computing community and fills a niche not addressed by existing Python numerical libraries. NumPy and SciPy.sparse each support the creation and operation of matrices (dense in the former case and sparse in the latter), but while NumPy does have structured arrays which support named fields by way of composing labeled datatypes, this approach does not scale in such a way as to handle the large amounts of data encountered in practice. Another related Python library is xarray \cite{hoyer2017xarray}, which also handles arrays with labeled rows and columns. xarray supports some numerical array operations, especially those which aggregate elements or which are applied entry-wise. Despite these similarities D4M.py and xarray differ substantially in intended usage, 
% -- by way of analogy, xarray is to dataframes what D4M is to sparse matrices. This difference in focus is illustrated well by the meaning and intention of $\sf{\_\_matmul\_\_}$: 
e.g., in xarray $\sf{A} \matmul \sf{B}$ is the dot product of compatible xarray data arrays, while in D4M $\sf{A} \matmul \sf{B}$ is the associative array product of two associative arrays, generalizing matrix multiplication. 

D4M.py implements all foundational functionality of D4M, including Accumulo and SQL database interface functionality via the Graphulo library. Graphulo implements matrix math primitives and graph algorithm building blocks in the style of GraphBLAS on top of Accumulo, representing database tables as D4M associative arrays and allowing D4M.py to operate on a massive scale using a highly scalable sorted, distributed key/value store. \cite{Graphulo2015,Graphulo2022}

\subsection{Mathematical Background}

D4M associative arrays are based on the mathematical notion of associative arrays, which generalize matrices. An $m \times n$ matrix $\ma{A}$, with $m$ and $n$ positive integers, is a rectangular array of (often) real numbers $a_{ij}$ ($1 \leq i \leq m$ and $1 \leq j \leq n$). \cite{Cayley1858}
\begin{equation*}
	\bf{A} = \begin{bmatrix} a_{11} & a_{12} & \cdots & a_{1n} \\ a_{21} & a_{22} & \cdots & a_{2n} \\ \vdots & \vdots & \ddots & \vdots \\ a_{m1} & a_{m2} & \cdots & a_{mn} \end{bmatrix}
\end{equation*}
One way to describe this more formally is as a function
\begin{equation*}
	\mathbf{A} \colon \{1, 2, \ldots, m\} \times \{1, 2, \ldots, n\} \to \mathbb{R},
\end{equation*}
with the quantity $\mathbf{A}(i, j)$ serving as the $(i, j)$-th entry $a_{ij}$ of the matrix $\ma{A}$ for each $1 \leq i \leq m$ and $1 \leq j \leq n$. 

Associative arrays generalize matrices in two ways: 
\begin{enumerate}[1.]
	\item The sets of row and column labels may be \emph{any} sets, not necessarily of the form $\{1, 2, \ldots, n\}$.
	\item The values may be drawn from instances of a general algebraic structure called a \textdef{semiring} supporting notions of addition and multiplication, not necessarily the real numbers $\mathbb{R}$ equipped with the standard notions of real number addition and multiplication.\footnote{More general definitions of matrices require that values are taken either from a field or a ring, algebraic structures which are generalized further by semirings.} 
\end{enumerate}
In other words, an associative array is a function of the form
\begin{equation*}
	\mathbf{A} \colon I \times J \to V,
\end{equation*}
where $I$ and $J$ are sets consisting of row keys and column keys, respectively; the value set $V$ is the underlying set of a semiring $(V, \oplus, \otimes, \se{0}, \se{1})$; and $\se{A}(i, j) = \se{0}$ for all but finitely many pairs $(i, j) \in I \times J$. \cite{kepner2018mathematics}

A \emph{semiring} is a mathematical structure with two binary operations $\oplus$ and $\otimes$ and two constant $\se{0}$ and $\se{1}$ subject to the following constraints for all $\se{u}, \se{v}, \se{w} \in V$: \cite{golan1999,GondranMinoux2007}
\begin{description}
	\item[Associativity of $\oplus$:] $\se{u} \oplus (\se{v} \oplus \se{w}) = (\se{u} \oplus \se{v}) \oplus \se{w}$.
	\item[Associativity of $\otimes$:] $\se{u} \otimes (\se{v} \otimes \se{w}) = (\se{u} \otimes \se{v}) \otimes \se{w}$.
	\item[Commutativity of $\oplus$:] $\se{u} \oplus \se{v} = \se{v} \oplus \se{u}$.
	\item[$\se{0}$ is identity for $\oplus$:] $\se{u} \oplus \se{0} = \se{u} = \se{0} \oplus \se{u}$.
	\item[$\se{1}$ is identity for $\otimes$:] $\se{u} \otimes \se{1} = \se{u} = \se{1} \otimes \se{u}$.
	\item[$\se{0}$ is annihilator for $\otimes$:] $\se{u} \otimes \se{0} = \se{0} = \se{u} \otimes \se{0}$.
	\item[Distributivity of $\otimes$ over $\oplus$:] $\se{u} \otimes (\se{v} \oplus \se{w}) = (\se{u} \otimes \se{v}) \oplus (\se{u} \otimes \se{w})$ and $(\se{v} \oplus \se{w}) \otimes \se{u} = (\se{v} \otimes \se{u}) \oplus (\se{w} \otimes \se{u})$.
\end{description}
Variations of semirings include \textdef{commutative semirings} (which additionally require that $\otimes$ be commutative) and \textdef{nonunital semirings} (which drop the necessity of an identity $\se{1}$ for $\otimes$). 
Common semirings include:
\begin{itemize}
	\item The plus-times algebra $(\mathbb{R}, +, \times, 0, 1)$.
	\item %$(\mathbb{R} \cup \{\pm \infty\}, \min, +, \infty, 0)$, the min-plus algebra, and 
	The max-plus algebra $(\mathbb{R} \cup \{\pm \infty\}, \max, +, -\infty, 0)$.
	\item The max-min algebra $(\mathbb{R} \cup \{\pm \infty\}, \max, \min, -\infty, \infty)$.
	\item The (nonunital) string algebra $(\Sigma^\star, \concat, \min, \epsilon)$ where $\Sigma$ is some totally ordered alphabet, $\Sigma^\star$ is the set of finite strings over the alphabet $\Sigma$, $\concat$ is string concatenation, $\min$ is the binary minimum taken with respect to the dictionary ordering induced by the implicit total ordering on $\Sigma$, and $\epsilon$ denotes the empty string.
%	\item Any bounded distributive lattice $(L, \vee, \wedge, 0, 1)$ where $0 = \min(L)$ and $1 = \max(L)$, including the max-min algebra $(\mathbb{R} \cup \{\pm \infty\}, \max, \min, -\infty, \infty)$.
\end{itemize}

Based on the first of the two defining characteristics of associative arrays is the allowance of arbitrary row and column keys: given an associative array $\ma{A} \colon I \times J \to V$ and two arbitrary indices $i, j$, then we make the convention that 
\begin{equation*}
	\ma{A}(i, j) = \se{0} \text{ whenever } (i, j) \notin I \times J,
\end{equation*}
as if by `padding' $\ma{A}$ with additional zero entries. 

The requirement that the values of an associative array fall within a set supporting sufficiently well-behaved notions of addition and multiplication allows associative arrays to support their own versions of matrix multiplication, matrix (element-wise) addition, and matrix element-wise multiplication. Let $\mathbf{A} \colon I_1 \times J_1 \to V$ and $\mathbf{B} \colon I_2 \times J_2 \to V$ be associative arrays.
\begin{description}
	\item[Associative Array Multiplication:] The product 
	\begin{equation*}
	    \mathbf{C} \coloneq \mathbf{A} \arrayprod \mathbf{B} \colon I_1 \times J_2 \to V    
	\end{equation*}
	is defined for each $i \in I_1$ and $j \in J_2$ by
	\begin{equation*}
		\mathbf{C}(i, j) \coloneq \bigoplus_{k \in J_1 \cap I_2}{(\mathbf{A}(i, k) \otimes \mathbf{B}(k, j))}.
	\end{equation*}
	
	Note that the well-definedness of this definition depends on the associativity and commutativity of $\oplus$ as well as the condition that $\mathbf{A}$ and $\mathbf{B}$ are nonzero for only finitely many pairs of keys. 
	
	\item[Associative Array Element-Wise Addition:] The sum 
	\begin{equation*}
	    \mathbf{C} \coloneq \mathbf{A} \oplus \mathbf{B} \colon (I_1 \cup I_2) \times (J_1 \cup J_2) \to V    
	\end{equation*} 
	is defined for each $i \in I_1 \cup I_2$ and $j \in J_1 \cup J_2$ by
	\begin{equation*}
		\mathbf{C}(i, j) \coloneq \mathbf{A}(i, j) \oplus \mathbf{B}(i, j).
%		\mathbf{C}(i, j) \coloneq \begin{cases} \mathbf{A}(i, j) \oplus \mathbf{B}(i, j) & \text{if $(i, j) \in (I_1 \times J_1) \cap (I_2 \times J_2)$,} \\ \mathbf{A}(i, j) & \text{if $(i, j) \in (I_1 \times J_1) \setminus (I_2 \times J_2)$,} \\ \mathbf{B}(i, j) & \text{if $(i, j) \in (I_2 \times J_2) \setminus (I_1 \times J_1)$,} \\ \se{0} & \text{otherwise.} \end{cases}
	\end{equation*}
	
	\item[Associative Array Element-Wise Multiplication:] The element-wise product 
	\begin{equation*}
	    \mathbf{C} \coloneq \mathbf{A} \otimes \mathbf{B} \colon (I_1 \cap I_2) \times (J_1 \cap J_2) \to V
	\end{equation*}
	is defined by for each $i \in I_1 \cap I_2$ and $j \in J_1 \cap J_2$ by
	\begin{equation*}
		\mathbf{C}(i, j) \coloneq \mathbf{A}(i, j) \otimes \mathbf{B}(i, j).
	\end{equation*}
\end{description}
The algebraic properties of semiring addition and multiplication imply that associative array multiplication, element-wise addition, and element-wise multiplication are associative, that associative array element-wise addition is commutative, and that both associative array multiplication and element-wise multiplication distribute over associative array element-wise addition.

\subsection{D4M Associative Arrays as Mathematical Associative Arrays}

In practice, both row and column key spaces for D4M associative arrays are assumed to consist of all strings and numbers, while the value set is either\ldots
\begin{description}
	\item[\ldots] the set of all numbers -- supporting the arithmetic operations $+, -, \times, /$ and the order-theoretic operations of $\min$ and $\max$ induced by the usual ordering of real numbers -- or\ldots
	\item[\ldots] the set of all strings -- supporting concatenation and the order-theoretic operations of $\min$ and $\max$ induced by the dictionary ordering.
\end{description}
D4M associative array values are assumed to either be entirely numerical or entirely strings, with the implicit semirings being the plus-times algebra $(\mathbb{R}, +, \times, 0, 1)$ and the (nonunital) string algebra $(\Sigma^\star, \concat, \min, \epsilon)$, respectively. We use the term ``nonempty'' instead of ``nonzero'' or ``nonnull'' to reflect the fact that ``zero'' may either be the integer $0$, the float $0.0$, the empty string $\epsilon$, etc., and reflects the fact that ``zeroes'' are unstored and contribute an empty space in visual representations of sparse arrays. 

As the full key space for D4M associative arrays is implicitly known, to construct an associative array $\sf{A}$ it suffices to know only the triples $(i, j, v)$ for which $\mathsf{A}[i, j] = v$ with $v \neq \se{0}$ (i.e., nonempty). 

To distinguish abstract associative arrays from D4M associative arrays we use bold font for the former, as in 
\begin{equation*}
    \mathbf{A} \colon I \times J \to V,
\end{equation*} 
and sans serif font for the latter and the Python language more generally, as in 
\begin{equation*}
    \mathsf{A} \coloneq \mathsf{D4M.assoc.Assoc(row, col, val)}. 
\end{equation*}
We occasionally mix mathematical notation and the Python language when there is no possibility of confusion and doing so helps with communication. E.g., in both contexts ``$\coloneq$'' is used to denote assignment while ``$=$'' is used to denote logical equality; ``$\leq$" is used in both contexts; etc. 

D4M.py follows the notational conventions of NumPy, using ``$@$'' to denote associative array multiplication, ``$+$'' to denote associative array element-wise addition, and ``$\ast$'' to denote associative array element-wise multiplication.

\section{Approach}

\subsection{D4M.py Associative Arrays}

In both D4M-MATLAB and D4M.jl, an associative array $\sf{A}$ is stored via four attributes, the first two of which are sorted one-dimensional arrays $\sf{A.row}$ and $\sf{A.col}$ containing the unique row and column keys associated with nonempty entries of the associative array, respectively. The remaining two attributes' particulars depend upon whether the values are numerical or not: In the numerical case, the third attribute $\sf{A.val}$ is simply the floating-point number $1.0$ (serving as a flag that the values are numerical) while the fourth attribute $\sf{A.adj}$ is a two-dimensional sparse matrix for which there is a one-to-one correspondence
\begin{equation*}
	\sf\mathsf{A}[\mathsf{A.row}[i], \mathsf{A.col}[j]] = v \iff \mathsf{A.adj}[i, j] = v.
\end{equation*}
In other words, in the numerical case values are stored directly within $\sf{A.adj}$.
In the string case, the third attribute $\sf{A.val}$ is the sorted one-dimensional array of all unique nonempty values and the fourth attribute $\sf{A.adj}$ is a two-dimensional sparse matrix for which there is a one-to-one correspondence
\begin{equation*}
	\sf\mathsf{A}[\mathsf{A.row}[i], \mathsf{A.col}[j]] = \mathsf{A.val}[k] \iff \mathsf{A.adj}[i, j] = k.\footnote{Note that both MATLAB and Julia are 1-indexed.}
\end{equation*}
In other words, in the string case the sparse matrix $\sf{A.adj}$ contains pointers to the values stored in $\sf{A.val}$. In both cases the arrays $\sf{A.row}$ and $\sf{A.col}$ serve to label the rows and columns of $\sf{A.adj}$, respectively. 

We continue this approach in D4M.py, with an associative array $\sf{A}$ stored via the following four attributes:
\begin{description}
	\item[$\sf{A.row}$:] A one-dimensional NumPy array of sorted, unique row keys associated with the nonempty entries of $\sf{A}$.
	\item[$\sf{A.col}$:] A one-dimensional NumPy array of sorted, unique column keys associated with the nonempty entries of $\sf{A}$.
	\item[$\sf{A.val}$:] Either the 64-bit floating-point number $1.0$ (in the case of numerical data) or else a one-dimensional NumPy array of sorted, unique nonempty values.
	\item[$\sf{A.adj}$:] A two-dimensional SciPy.sparse matrix in COO (COOrdinate) format of size $\sf{len(A.row)} \times \sf{len(A.col)}$ such that when $\sf{not\ isinstance(A.val, float)}$, then the nonempty values of $\sf{A.adj}$ are exactly the integers from $1$ to $\sf{len(A.val)}$.
\end{description}
Because $\sf{A.adj}$ is a sparse matrix and Python is $0$-indexed, the correspondence for the nonnumerical case is:
\begin{equation*}
	\sf{A}[\sf{A.row}[i], \sf{A.col}[j]] = \sf{A.val}[k] \iff \sf{A.adj}[i, j] = k+1.
\end{equation*}
Figure~\ref{associative array data model example} describes these four attributes for the associative array depicted in Figure~\ref{associative array example}.

\begin{figure}[htb]
	\begin{empheq}[box=\fbox]{align*}
		\sf{A.row} & = \sf{np.array([``0294.mp3\text{''}, ``1829.mp3\text{''},} \\
		& \hphantom{= \sf{np.array([~}} \sf{``7802.mp3\text{''}])} \\
		\sf{A.col} & = \sf{np.array([``artist\text{''}, ``duration\text{''}, ``genre\text{''}])} \\
		\sf{A.val} & = \sf{np.array([``10\colon12\text{''}, ``6\colon53\text{''}, ``8\colon01\text{''},} \\
		& \hphantom{= \sf{np.array([~}} \sf{``Pink\ Floyd\text{''}, ``Samuel\ Barber\text{''},} \\
		& \hphantom{= \sf{np.array([~}} \sf{ ``Taylor\ Swift\text{''},} \\
		& \hphantom{= \sf{np.array([~}} \sf{``classical\text{''}, ``pop\text{''}, ``rock\text{''}])} \\
		\sf{A.adj} & = \sf{sp.coo\_matrix([[3, 1, 8], [4, 2, 6], [5, 0, 7]])}
	\end{empheq}
	\caption{D4M.py data model for the associative array in Figure~\ref{associative array example}. ``$\sf{np}$'' is an abbreviation for ``$\sf{numpy}$'' and ``$\sf{sp}$'' is an abbreviation for ``$\sf{scipy.sparse}$''.}
	\label{associative array data model example}
\end{figure}

An edge case exists in the empty associative array, which may either be considered numerical or string. In this case we store the associative array as if it numerical and must account for the possibility of emptiness when the distinction between numerical and string associative arrays is necessary.

There are two ways in which D4M.py associative arrays may be constructed:
\begin{enumerate}[1.]
	\item $\sf{A} \coloneq \sf{D4M.assoc.Assoc(row, col, val, aggregate{=}bin\_op)}$, where $\sf{row}$, $\sf{col}$, and $\sf{val}$ are sequences of strings or numbers each of which are the same length (or can be unambiguously broadcast to the same length). The optional $\sf{aggregate}$ parameter $\sf{bin\_op}$ indicates an associative, commutative binary operation (default $\sf{min}$) used to handle any collisions, i.e., instances where there are two indices $\sf{k\_1}, \sf{k\_2}$ for which $\sf{row}[k\_1] = \mathsf{row}[k\_2]$ and $\sf{col}[k\_1] = \mathsf{col}[k\_2]$. 
	
	\item $\sf{A} \coloneq \sf{D4M.assoc.Assoc(row, col, val, adj{=}sp\_mat)}$, where $\sf{row}$, $\sf{col}$, and $\sf{val}$ are sequences of strings or numbers and $\sf{sp\_mat}$ is a SciPy.sparse matrix. The sorted unique entries of $\sf{row}$ and $\sf{col}$ are extracted and cut down to the dimensions of $\sf{sp\_mat}$; if $\sf{isinstance(val, float)}$ then $\sf{A}$ is treated as a numerical associative array, otherwise the nonempty entries of $\sf{sp\_mat}$ are treated as the ($1$-indexed) indices of the sorted unique values in $\sf{val}$.
\end{enumerate}
%[Allowable constructor parameters.]

\subsection{Extraction \& Assignment}
%\subsection{\texorpdfstring{$\_\_\sf{getitem}\_\_$}{getitem} and \texorpdfstring{$\_\_\sf{setitem}\_\_$}{setitem}}

D4M.py associative arrays support both the $\sf{\_\_getitem\_\_}$ and $\sf{\_\_setitem\_\_}$ magic methods, but two ambiguities/subtleties that require additional comment exist for the former.
\begin{enumerate}[1.]
	\item Suppose $\sf{A}$ is an associative array whose row and column keys are strings. When extracting a subarray from $\sf{A}$ there are analogs for Python slice objects available, e.g., $\sf{``a,:,b\text{,''}}$ corresponds to the range of all keys $\sf k$ with $\sf{``a\text{''}} \leq k \leq \sf{``b\text{''}}$. In particular, these `string slices' are \emph{inclusive} on the right, unlike Python ranges or slices which are \emph{exclusive} on the right.
	
	\item In an expression like ``$\sf{A[1, 0 {\colon} 2]}$'' it is ambiguous whether the integers $0, 1, 2$ should be considered as indices of $\sf{A.row}$ and $\sf{A.col}$ or if they should be treated as members of $\sf{A.row}$ and $\sf{A.col}$. The row and column keys are most often strings, so the interpretation of slices and integer arrays as indices of $\sf{A.row}$ and $\sf{A.col}$ is taken.
\end{enumerate}

\subsection{Associative Array Algebra Operations}

The format in which the D4M.py associative arrays are stored -- the $\sf{.row}$, $\sf{.col}$, $\sf{.val}$, and $\sf{.adj}$ attributes -- is well-suited to the associative array algebra operations. The approach across D4M-MATLAB, D4M.jl, and now D4M.py is to leave as much of the work to a dedicated sparse linear algebra library -- MATLAB's in-built sparse linear algebra, Julia's SparseArrays stdlib module, and SciPy.sparse for Python. Let $\sf{A}$ and $\sf{B}$ denote D4M.py associative arrays.

\subsubsection{Associative Array Element-Wise Addition}

In the string case the sum $\sf{A} + \sf{B}$ is calculated by first extracting the triples $(\sf{row\_A}, \sf{col\_A}, \sf{val\_A})$ and $(\sf{row\_B}, \sf{col\_B}, \sf{val\_B})$ which can be used to reconstruct $\sf{A}$ and $\sf{B}$, respectively. Appending $\sf{row\_A}$ and $\sf{row\_B}$, $\sf{col\_A}$ and $\sf{col\_B}$, and $\sf{val\_A}$ and $\sf{val\_B}$ produces NumPy arrays $\sf{row\_C}$, $\sf{col\_C}$, and $\sf{val\_C}$ which can be used to construct the sum $\sf{A} + \sf{B}$ using concatenation as an aggregation function. Any collisions which take place occur between a value from $\sf{A}$ and a value from $\sf{B}$ and occur at most once for each pair of row and column keys. This approach is encapsulated in (and generalized by) the method called $\sf{D4M.assoc.Assoc.combine}$ which handles this string addition operation and the element-wise minimum and maximum operations. 

The numerical case allows for higher efficiency and makes use of an operation called sorted union. For repetition-free sorted iterables $\sf I$ and $\sf J$, the sorted union $\sf I \cup J$ is found by simultaneously iterating through $\sf I$ and $\sf J$ in an alternating fashion, appending to a new iterable $\sf K$ all encountered elements along the way. In further detail, suppose at some stage in the procedure the $\sf m$-th element of $\sf I$ and the $\sf n$-th element of $\sf J$ have been reached.
\begin{description}
	\item[Case 1:] $\sf I[m] > J[n]$. Repeatedly increment $\sf n$ until $\sf I[m] \leq J[n]$, appending elements of $\sf J$ to $\sf K$ along the way.
	\item[Case 2:] $\sf I[m] = J[n]$. Append this common element to $\sf K$ and increment $\sf m$ and $\sf n$.
	\item[Case 3:] $\sf I[m] < J[n]$. Analogous to Case 1 above, with the roles of $\sf I, m$ and $\sf J, n$ switched.
\end{description}
If at any point one or both of $\sf I$ and $\sf J$ are exhausted, the remaining elements are appended to $\sf K$, completing the procedure. During the construction of $\sf K$, index maps describing how $\sf I$ and $\sf J$ sit within $\sf K$ can be easily constructed.

Making use of the concurrently constructed index maps, $\sf{A.adj}$ and $\sf{B.adj}$ may be re-shaped and re-indexed to $(\sf{A.row} \cup \sf{B.row}) \times (\sf{A.col} \cup \sf{B.col})$).
%, as in
%\begin{equation*}
%	\sf{scipy.sparse.coo\_matrix((A.adj.data, (row\_map\_A[A.adj.row], col\_map\_A[A.adj.col])), shape=(len(row\_union), len(col\_union)), dtype=float).tocsr()},
%\end{equation*}
%where 
%\begin{align*}
%	\sf{row\_union}, \sf{row\_map_A}, \sf{row\_map\_B} & = \sf{sorted\_union(A.row, B.row, return\_index=True)} \\
%	\sf{col\_union}, \sf{col\_map\_A}, \sf{col\_map\_B} & = \sf{sorted\_union(A.col, B.col, return\_index=True)}.
%\end{align*}
The resulting sparse matrices may then be added directly using the built-in $\sf{scipy.sparse.csr\_matrix.\_\_add\_\_}$ to obtain a sparse matrix $\sf C\_adj\_pre$. 

Finally, a method called $\sf{D4M.assoc.Assoc.condense}$ is used to remove any resulting empty rows and columns; this is done by converting $\sf C\_adj\_pre$ to both the CSR (Compressed Sparse Row) and CSC (Compressed Sparse Column) formats, where the attributes $\sf csr\_rows \coloneq C\_adj\_pre.tocsr().indptr$ and $\sf csc\_cols \coloneq C\_adj\_pre.tocsc().indptr$ are extracted. The Boolean-typed NumPy arrays 
\begin{equation*}
    \sf good\_rows \coloneq (csr\_rows[{:}{-}1] < csr\_rows[1{:}])
\end{equation*} 
and 
\begin{equation*}
    \sf good\_cols \coloneq \sf (csc\_cols[{:}{-}1] < csc\_cols[1{:}])    
\end{equation*}
provide the indices for nonempty rows and columns, respectively. $\sf{C} \coloneq \sf{A} + \sf{B}$ is given by 
\begin{align*}
    \sf C.row & \coloneq \sf row\_union[good\_rows], \\
    \sf C.col & \coloneq \sf row\_union[good\_cols], \\
    \sf C.val & \coloneq \sf 1.0, \\
    \sf C.adj & \coloneq \sf C\_adj\_pre.tocsr()[good\_rows{:}][{:}good\_cols].tocoo().
\end{align*}

\subsubsection{Associative Array Element-Wise Multiplication}

Associative array element-wise multiplication makes use of a related operation on repetition-free sorted iterables called sorted intersection. Given two repetition-free sorted iterables $\sf I$ and $\sf J$ their sorted intersection $\sf K \coloneq I \cap J$ is formed in a way close to that of the sorted union $\sf I \cup J$ except that elements are appended to $\sf K$ only when corresponding to the case $\sf I[m] = J[n]$. While the sorted intersection is carried out, index maps are constructed which describe how $\sf K$ sits within $\sf I$ and $\sf J$. 

For the numerical case of associative array element-wise multiplication, the index maps constructed when computing the sorted intersections $\sf{A.row} \cap \sf{B.row}$ and $\sf{A.col} \cap \sf{B.col}$ are used to restrict and re-index $\sf{A.adj}$ and $\sf{B.adj}$ to $(\sf{A.row} \cap \sf{B.row}) \times (\sf{A.col} \cap \sf{B.col})$. The two sparse matrices may then be converted to the CSR format and element-wise multiplied using the $\sf{scipy.sparse.csr\_matrix.multiply}$ method. Dropping any empty rows and columns via the $\sf .condense()$ method produces $\mathsf{A} \ast \mathsf{B}$.

The string case is similar to that of associative array addition, making use of the fact that the default aggregation function is $\min$. Associative array element-wise multiplication allows for the mixed type case of a string associative array element-wise multiplied by a numerical associative array, with the latter acting as a mask on the former. Numerical associative array element-wise multiplication by a string associative array is also supported, but differs in its result, being reduced to the numerical case by calling the $\sf{B.logical()}$ method, which replaces each nonempty entry of $\sf{B}$ with $1$ -- this can be very easily achieved by replacing $\sf{B.val}$ with $1.0$ and $\sf{B.adj.data}$ with $\sf{np.ones(len(B.adj.data))}$.

\subsubsection{Associative Array Multiplication}

Sorted intersection is also utilized by associative array multiplication, arising in the sorted intersection $\sf{A.col} \cap \sf{B.row}$. The concurrently constructed index maps are used to restrict and re-index $\sf{A.adj}$ to $\sf{A.row} \times (\sf{A.col} \cap \sf{B.row})$ and $\sf{B.adj}$ to $(\sf{A.col} \cap \sf{B.row}) \times \sf{B.col}$. After converting to the CSR format the resulting sparse matrices can be multiplied using their native matrix multiplication. Using the $\sf .condense()$ method yields the associative array product $\sf{A} \matmul \sf{B}$.

Associative array multiplication is currently defined only for numerical associative arrays, so string associative arrays are converted via the $\sf{.logical()}$ method prior.

\section{Performance}

\subsection{Setup}

Benchmarking for the associative array constructor and binary associative array operations was done using D4M associative arrays of dimensions roughly $2^n \times 2^n$, $n$ ranging over $5 \leq n \leq 18$. The data used to generate the requisite associative arrays are stored within six \texttt{.txt} files, \texttt{rows.txt}, \texttt{rows2.txt}, \texttt{cols.txt}, \texttt{cols2.txt}, \texttt{num\_vals.txt}, and \texttt{string\_vals.txt}. Each consist of $14$ arrays, one for each $5 \leq n \leq 18$ and consisting of $8 \times 2^n$ elements -- uniformly random integers (cast as strings) between $0$ and $2^n$ for the first four files, uniformly random integers between $0$ and $100$ for the fifth file, and uniformly random strings of length $8$ for the sixth and final file. The files are loaded as lists of NumPy arrays in the variables $\mathsf{rows}$, $\mathsf{rows2}$, $\mathsf{cols}$, $\mathsf{cols2}$, $\mathsf{num\_vals}$, and $\mathsf{str\_vals}$, respectively. 

Five benchmarking tests are run:
\begin{enumerate}[1.] 
	\item  $\sf \mathsf{D4M.assoc.Assoc}(\mathsf{rows}[n], \mathsf{cols}[n], \mathsf{num\_vals}[n])$.
	\item $\sf \mathsf{D4M.assoc.Assoc}(\mathsf{rows}[n], \mathsf{cols}[n], \mathsf{str\_vals}[n])$. 
\end{enumerate}
With $\sf{A} \coloneq \mathsf{D4M.assoc.Assoc}(\mathsf{rows}[n], \mathsf{cols}[n], \mathsf{1})$ and $\sf{B} \coloneq \mathsf{D4M.assoc.Assoc}(\mathsf{rows2}[n], \mathsf{cols2}[n], \mathsf{1})$\ldots
\begin{enumerate}[1., resume]
	\item $\sf{A} + \sf{B} ~~ (= \sf{A.\_\_add\_\_(B)})$.
	\item $\sf{A} \matmul \sf{B} ~~ (= \sf{A.\_\_matmul\_\_(B)})$.
	\item $\sf{A} \ast \sf{B} ~~ (= \sf{A.\_\_mul\_\_(B)})$.
%	\item $\mathsf{D4M.assoc.Assoc}(\mathsf{rows}[n], \mathsf{cols}[n], \mathsf{1}) + \mathsf{D4M.assoc.Assoc}(\mathsf{rows2}[n], \mathsf{cols2}[n], \mathsf{1})$.
%	\item $\mathsf{D4M.assoc.Assoc}(\mathsf{rows}[n], \mathsf{cols}[n], \mathsf{1}) \matmul \mathsf{D4M.assoc.Assoc}(\mathsf{rows2}[n], \mathsf{cols2}[n], \mathsf{1})$.
%	\item $\mathsf{D4M.assoc.Assoc}(\mathsf{rows}[n], \mathsf{cols}[n], \mathsf{1}) \ast \mathsf{D4M.assoc.Assoc}(\mathsf{rows2}[n], \mathsf{cols2}[n], \mathsf{1})$.
\end{enumerate}

Our tests were run on a single Intel Xeon-P8 core on the MIT SuperCloud system, recording running time in seconds which was averaged over $10$ runs.

%We ran our tests on two platforms. The first platform is a laptop with specifications of an i7-8650U CPU (1.90~GHz, 2112~Mhz, 4 Core(s), 8 Logical Processor(s)

\subsection{Measurements}

\begin{figure}[htb]
	\centering
	\includegraphics[scale=.6]{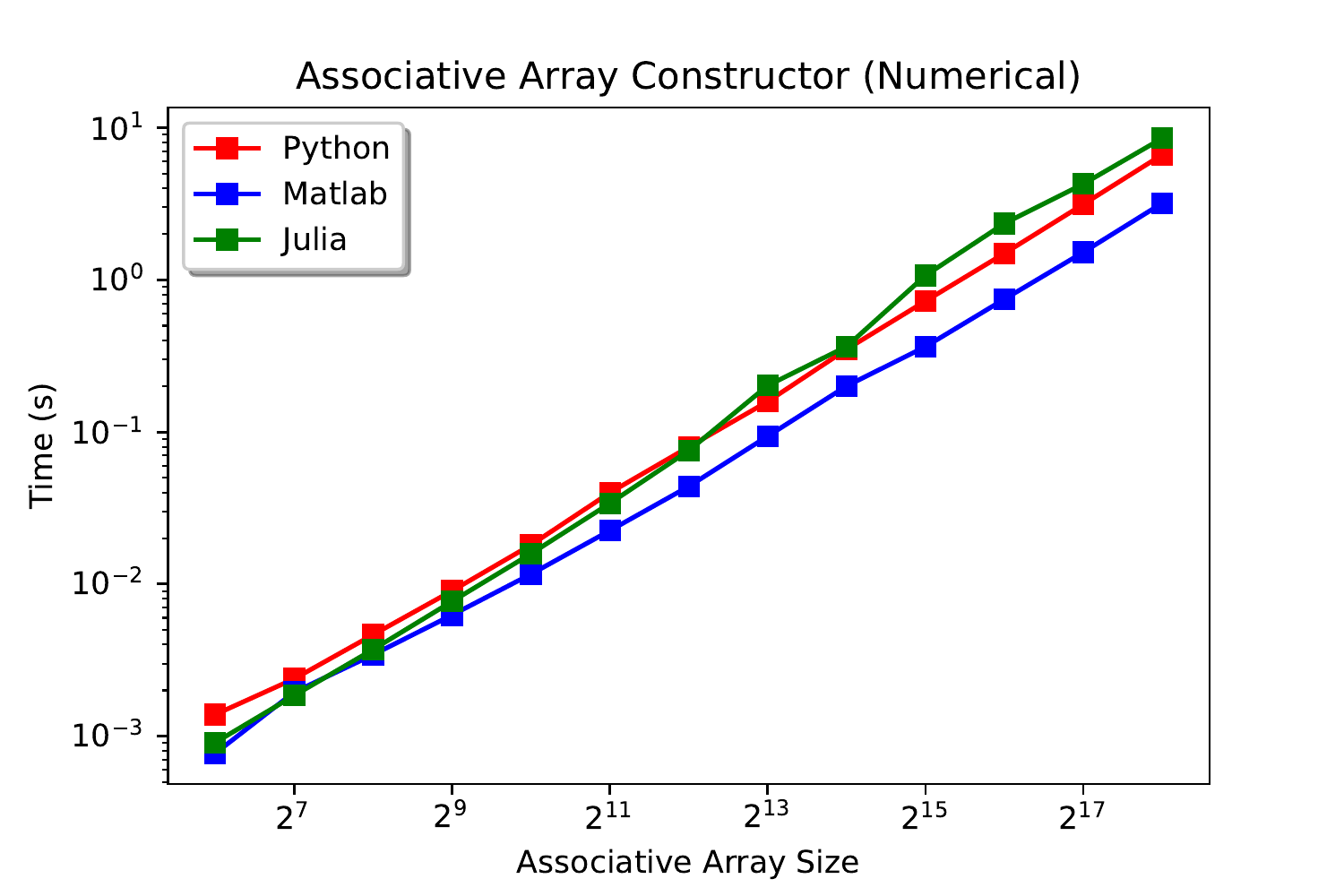}
	\caption{Runtime of $\sf{D4M.assoc.Assoc}$ Constructor in the case of numerical values across associative arrays of dimensions $2^n \times 2^n$ with $\approx 8$ integer entries between $0$ and $100$ per row chosen uniformly at random, where $5 \leq n \leq 18$.}
	\label{assoc constructor num figure}
\end{figure}

\begin{figure}[htb]
	\centering
	\includegraphics[scale=.6]{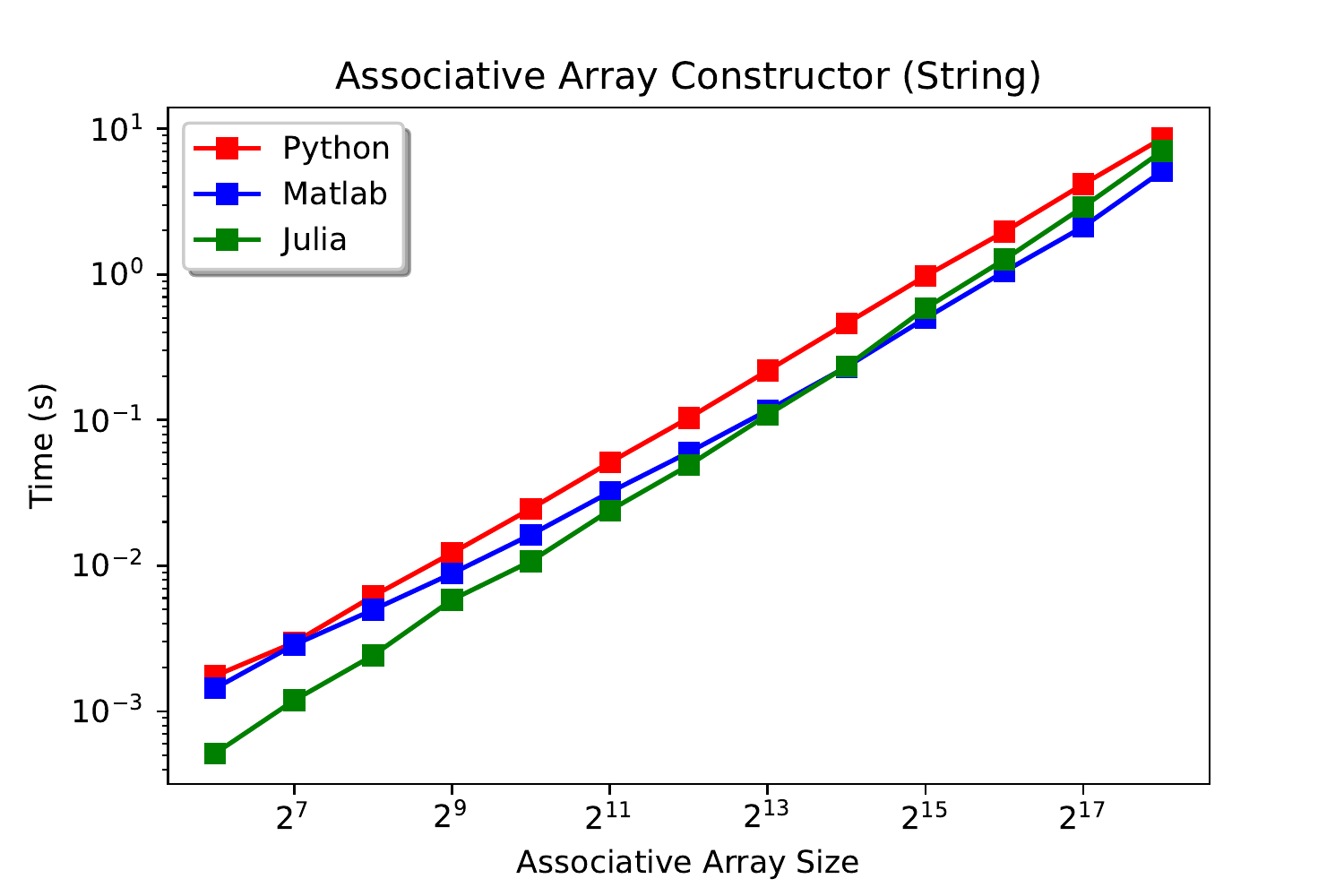}
	\caption{Runtime of $\sf{D4M.assoc.Assoc}$ Constructor in the case of string values across associative arrays of dimensions $2^n \times 2^n$ with $\approx 8$ string entries of length $8$ per row chosen uniformly at random, where $5 \leq n \leq 18$.}
	\label{assoc constructor str figure}
\end{figure}

The first two tests benchmark the $\mathsf{D4M.assoc.Assoc}$ constructor both in the case where the $\mathsf{val}$ argument is a numerical array ($\sf{num\_vals}[n]$) and in the case where it is instead an array of strings ($\sf{str\_vals[n]}$). In both cases the $\mathsf{row}$ and $\mathsf{col}$ arguments are identical for each $n$, $\sf{rows}[n]$ and $\sf{cols}[n]$. Figure~\ref{assoc constructor num figure} shows the runtime of the associative array constructor with numerical values across Python, MATLAB, and Julia, showing comparable performance across all three. Figure~\ref{assoc constructor str figure} shows similar behavior with the three implementations having comparable performance as $n$ increases.

\begin{figure}[htb]
	\centering
	\includegraphics[scale=.6]{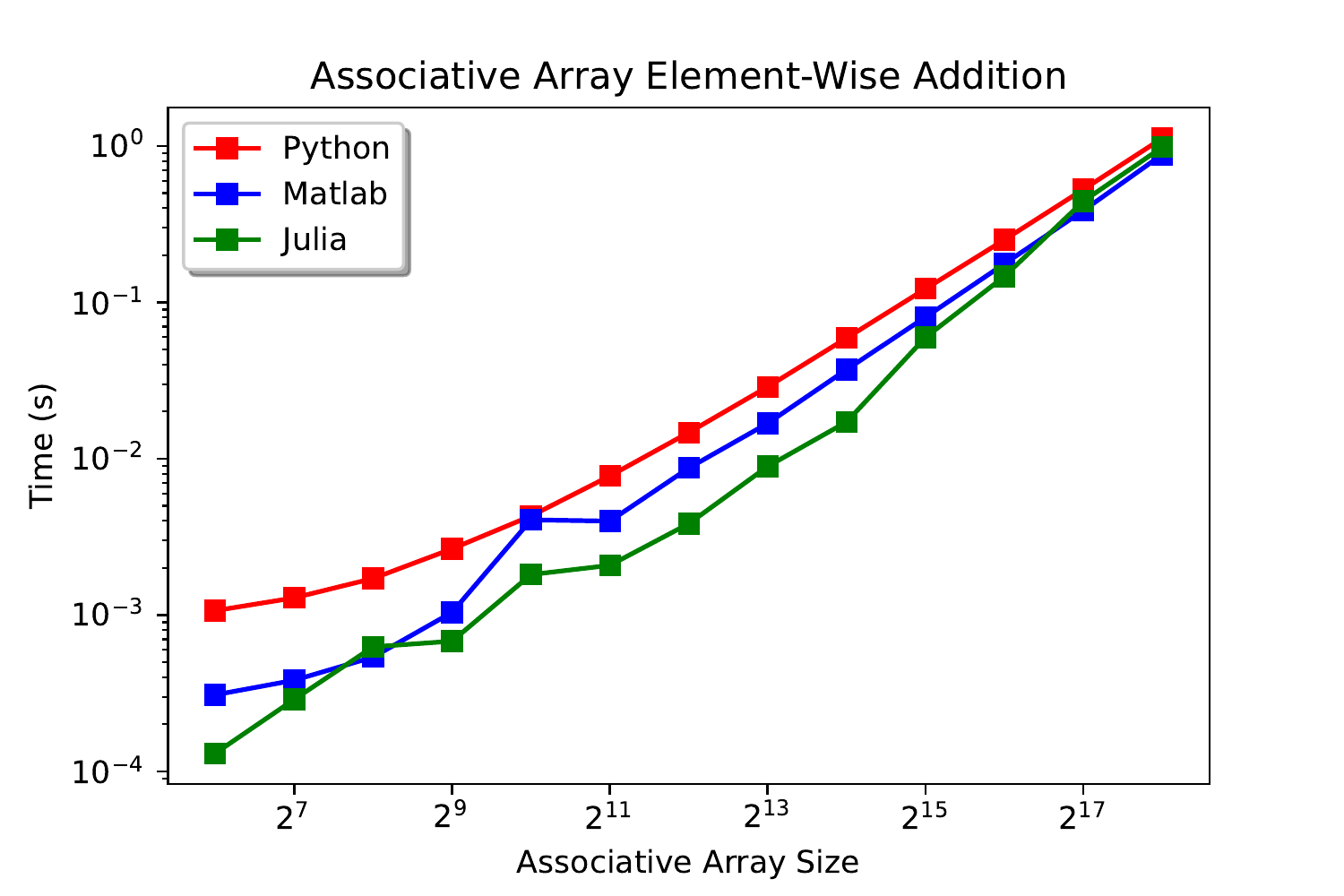}
	\caption{
	    Runtime of associative array element-wise addition, $\sf{D4M.assoc.Assoc.\_\_add\_\_}$, across associative arrays of dimensions $2^n \times 2^n$ with $\approx 8$ nonempty entries of value $1$ per row, where $5 \leq n \leq 18$.
	}
	\label{assoc addition figure}
\end{figure}

Figure~\ref{assoc addition figure} shows the results of the associative array element-wise addition benchmarking test. The three implementations show comparable performance, with the runtimes getting steadily closer as $n$ increases. 

\begin{figure}[htb]
	\centering
	\includegraphics[scale=.6]{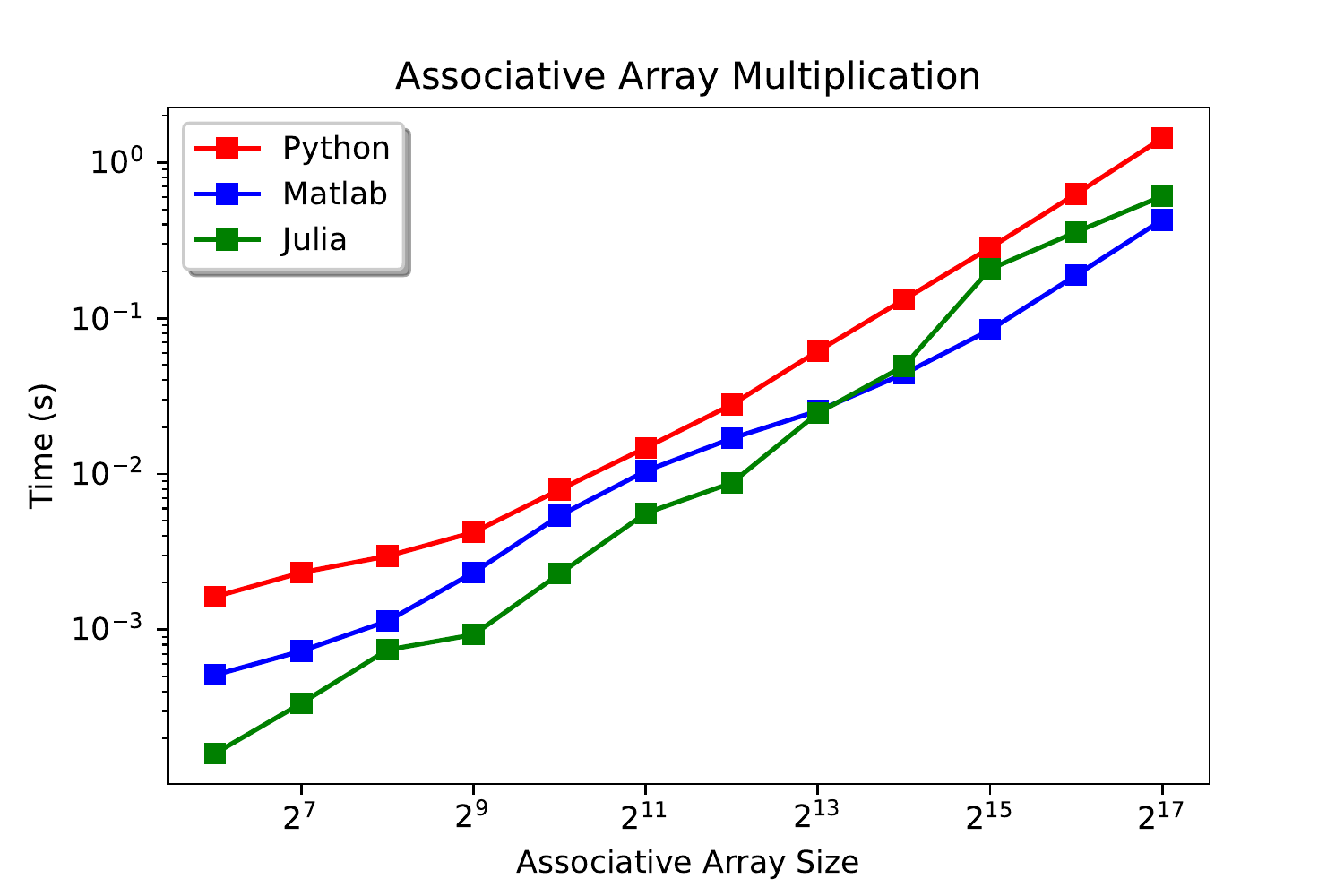}
	\caption{
	    Runtime of associative array multiplication, $\sf{D4M.assoc.Assoc.\_\_matmul\_\_}$, across associative arrays of dimensions $2^n \times 2^n$ with $\approx 8$ nonempty entries of value $1$ per row, where $5 \leq n \leq 17$.
	}
	\label{assoc multiplication figure}
\end{figure}

Figure~\ref{assoc multiplication figure} shows the results of the associative array multiplication benchmarking test. The three implementations show comparable performance as $n$ increases, within one order of relative magnitude of each other. Profiling D4M.py's $\mathsf{\_\_matmul\_\_}$ shows that two things dominate the running time: conversion between sparse matrix formats while carrying out the sparse matrix multiplication and the clean-up afterwards when removing empty rows and columns via the $\sf .condense()$ method. 

\begin{figure}[htb]
	\centering
	\includegraphics[scale=.6]{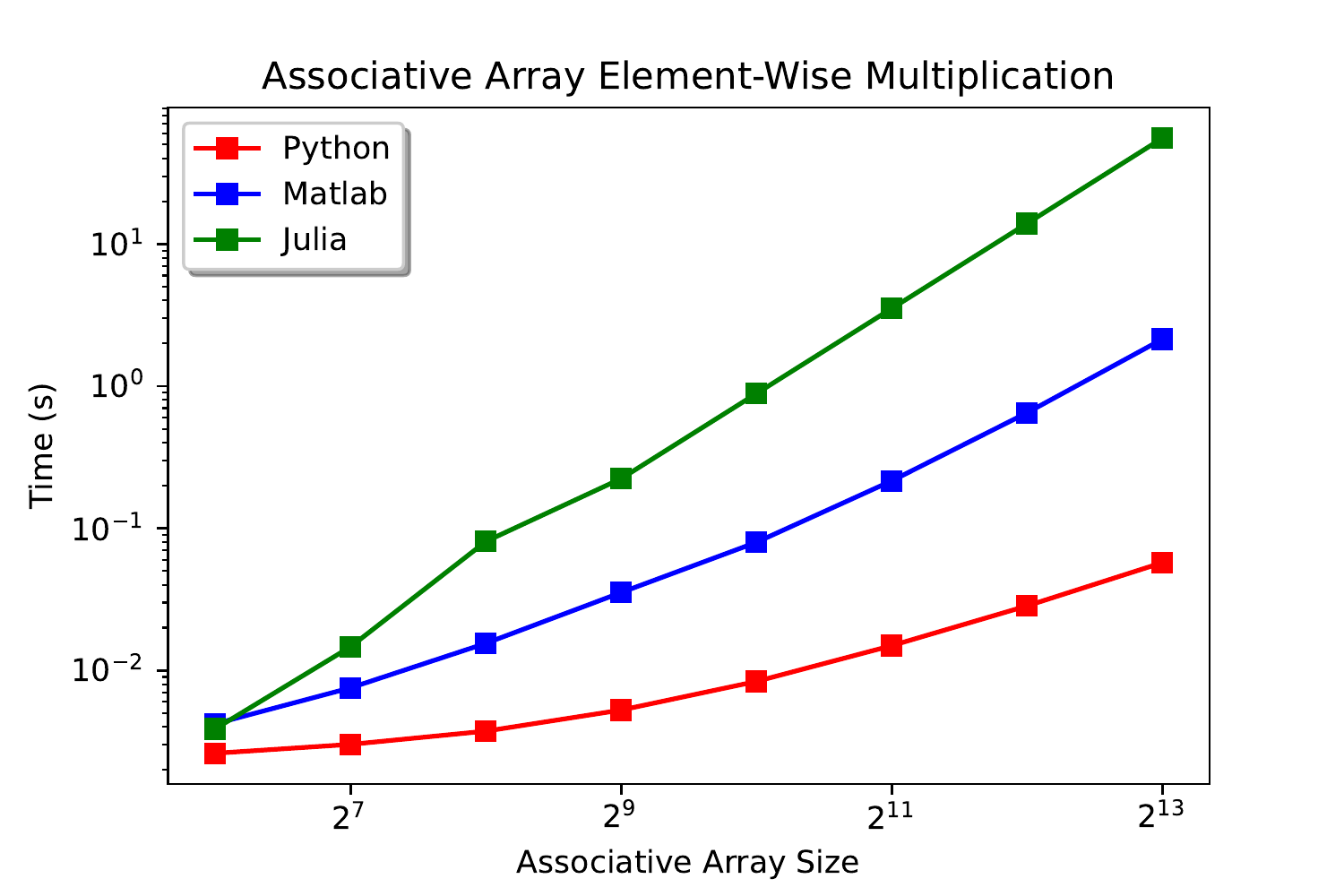}
	\caption{
	    Runtime of associative array element-wise multiplication, $\sf{D4M.assoc.Assoc.\_\_mul\_\_}$, across associative arrays of dimensions $2^n \times 2^n$ with $\approx 8$ nonempty entries of value $1$ per row, where $5 \leq n \leq 13$.
	}
	\label{assoc hadamard multiplication figure}
\end{figure}

Finally, Figure~\ref{assoc hadamard multiplication figure} depicts the results of the associative array element-wise multiplication benchmarking test. Unlike in the previous benchmarking tests, all three runtime curves diverge from one another, with Python remaining flattest among the three. The loop structure of associative array element-wise multiplication is very similar to that of associative array element-wise addition, so the performance curves should be expected to be similar. This indicates that the MATLAB and Julia implementations of element-wise multiplication can be optimized further. Because of the large running times relative to $n$, only $n$ up to $13$ were evaluated during the test.

\section{Conclusion \& Future Work}

D4M.py further enhances Python's role as a standard scientific computing language by providing a composable toolbox which can operate on structured or unstructured data, particularly geared towards applications concerning matrices, spreadsheets, databases, and graphs. The library complements existing Python numerical libraries like NumPy and SciPy and has the potential to complement other standard, new, or emerging Python libraries which deal with big data. Among the tested associative array functions between the three existing implementations of D4M, D4M.py's performance is either within an order of magnitude of the most efficient implementation or is the most efficient implementation itself (as in the case of the associative array element-wise product).

While D4M.py currently uses SciPy.sparse as its underlying sparse linear algebra library, we plan to distill and modularize the sparse linear algebra functions utilized so that other sparse linear algebra libraries can be easily used in lieu of SciPy.sparse. Such possibilities include newly emerging (or newly accessible) sparse linear algebra libraries like PyGraphBLAS \cite{Pelletier2021}, Python-GraphBLAS \cite{grblas2019}, and PyGB \cite{PyGB2018} which provide Python wrappers around the GraphBLAS API and interfaces to the highly performant SuiteSparse:GraphBLAS library \cite{Davis2019,Davis2022}, as well as software packages which plan to support sparse multidimensional arrays in the future like Arkouda \cite{arkouda2020,arkouda2021}. These developments could provide further opportunity to optimize existing D4M.py capabilities and extend them further. The notion of semirings and more general algebraic structures supporting a notion of ``addition'', ``multiplication'', and ``zero'' is also present in GraphBLAS, suggesting deeper support for user-selected or user-defined semiring operations within D4M. 

Historically D4M has provided an interface for key-value store and relational databases. Extending D4M.py's support for databases like Accumulo and SciDB can further lower the barrier to entry for data analysts and data scientists to utilize those databases, so we plan to further develop those capabilities.

\section*{Acknowledgment}

The authors wish to acknowledge the following individuals for their contributions and support: Bob Bond, Stephen Buckley, Tucker Hamilton, Jeff Gottschalk, Chris Hill, Tim Kraska, Charles Leiserson, Mimi McClure, Kyle McAlpin, Joseph McDonald, Sandy Pentland, Heidi Perry, Christian Prothmann, John Radovan, Steve Rejto, Daniela Rus, Matthew Weiss, Marc Zissman.

%The preferred spelling of the word ``acknowledgment'' in America is without 
%an ``e'' after the ``g''. Avoid the stilted expression ``one of us (R. B. 
%G.) thanks $\ldots$''. Instead, try ``R. B. G. thanks$\ldots$''. Put sponsor 
%acknowledgments in the unnumbered footnote on the first page.

%\section*{References}

%Please number citations consecutively within brackets \cite{b1}. The 
%sentence punctuation follows the bracket \cite{b2}. Refer simply to the reference 
%number, as in \cite{b3}---do not use ``Ref. \cite{b3}'' or ``reference \cite{b3}'' except at 
%the beginning of a sentence: ``Reference \cite{b3} was the first $\ldots$''
%
%Number footnotes separately in superscripts. Place the actual footnote at 
%the bottom of the column in which it was cited. Do not put footnotes in the 
%abstract or reference list. Use letters for table footnotes.
%
%Unless there are six authors or more give all authors' names; do not use 
%``et al.''. Papers that have not been published, even if they have been 
%submitted for publication, should be cited as ``unpublished'' \cite{b4}. Papers 
%that have been accepted for publication should be cited as ``in press'' \cite{b5}. 
%Capitalize only the first word in a paper title, except for proper nouns and 
%element symbols.
%
%For papers published in translation journals, please give the English 
%citation first, followed by the original foreign-language citation \cite{b6}.

\bibliographystyle{IEEEtran}
\bibliography{hpec_d4m}

%\begin{thebibliography}{00}
%\bibitem{D4M2012} 
%%\bibitem{b1} G. Eason, B. Noble, and I. N. Sneddon, ``On certain integrals of Lipschitz-Hankel type involving products of Bessel functions,'' Phil. Trans. Roy. Soc. London, vol. A247, pp. 529--551, April 1955.
%%\bibitem{b2} J. Clerk Maxwell, A Treatise on Electricity and Magnetism, 3rd ed., vol. 2. Oxford: Clarendon, 1892, pp.68--73.
%%\bibitem{b3} I. S. Jacobs and C. P. Bean, ``Fine particles, thin films and exchange anisotropy,'' in Magnetism, vol. III, G. T. Rado and H. Suhl, Eds. New York: Academic, 1963, pp. 271--350.
%%\bibitem{b4} K. Elissa, ``Title of paper if known,'' unpublished.
%%\bibitem{b5} R. Nicole, ``Title of paper with only first word capitalized,'' J. Name Stand. Abbrev., in press.
%%\bibitem{b6} Y. Yorozu, M. Hirano, K. Oka, and Y. Tagawa, ``Electron spectroscopy studies on magneto-optical media and plastic substrate interface,'' IEEE Transl. J. Magn. Japan, vol. 2, pp. 740--741, August 1987 [Digests 9th Annual Conf. Magnetics Japan, p. 301, 1982].
%%\bibitem{b7} M. Young, The Technical Writer's Handbook. Mill Valley, CA: University Science, 1989.
%\end{thebibliography}

\end{document}